\documentclass[preprint,authoryear,12pt]{elsarticle}

\usepackage{epsfig}

\usepackage{amssymb}

\journal{Advances in Space Research}

\begin{document}

%% Frontmatter
\begin{frontmatter}

%% Title, authors and addresses

% Use the tnoteref command within \title and fnref within \author or \address for footnotes;
% use the corref command within \author for corresponding author footnotes;
% use the ead command for the email address,
% and the form \ead[url] for the home page:
% \title{Title\tnoteref{label1}}
% \tnotetext[label1]{}
% \author{Name\corref{cor1}\fnref{label2}}
% \ead{email address}
% \ead[url]{home page}
% \fntext[label2]{}
% \cortext[cor1]{}
% \address{Address\fnref{label3}}
% \fntext[label3]{}

%%%%%%%%%%%%%%%%%%%%%%%%%%%%%%%%%%%%%%%%%%%%%%%%%%%%%%%%%%%%
%%%%%%%%%%%%%%%%%%%%%%%%%%%%%%%%%%%%%%%%%%%%%%%%%%%%%%%%%%%%

\title{Active galactic nuclei imaging programs of the \emph{RadioAstron} mission}%\tnoteref{footnote1}}
%\tnotetext[footnote1]{This template can be used for all publications in Advances in Space Research.}

% Use optional labels to link authors explicitly to addresses:
% \author[label1,label2]{}
% \address[label1]{}
% \address[label2]{}

%%%%%%%%%%%%%%%%%%%%%%%%%%%%%%%%%%%%%%%%%%%%%%%%%%%%%%%%%%%%
%%%%%%%%%%%%%%%%%%%%%%%%%%%%%%%%%%%%%%%%%%%%%%%%%%%%%%%%%%%%

\author[v1]{Gabriele Bruni\corref{cor}}
\address[v1]{INAF - Istituto di Astrofisica e Planetologia Spaziali, via del Fosso del Cavaliere 100, 00133 Rome, Italy}
\cortext[cor]{Corresponding author}
%\fntext[footnote2]{Additional information regarding the corresponding author}
\ead{gabriele.bruni@inaf.it}

% Url can be given like this:
% \ead[url]{http://www.elsevier.com/wps/find/authorsview.authors/latex}

%%%%%%%%%%%%%%%%%%%%%%%%%%%%%%%%%%%%%%%%%%%%%%%%%%%%%%%%%%%%

\author[i1,i2,i3]{Tuomas Savolainen}%\fnref{footnote1}}

\address[i1]{Aalto University Department of Electronics and Nanoengineering, PL 15500, FI-00076 Aalto, Finland}
\address[i2]{Aalto University Mets\"ahovi Radio Observatory, Mets\"ahovintie 114, 02540 Kylm\"al\"a, Finland}
\address[i3]{Max-Planck-Institut f\"ur Radioastronomie,  Auf dem H\"ugel 69, 53121 Bonn, Germany}
%\fntext[footnote1]{Additional affiliations: Aalto University Mets\"ahovi Radio Observatory, Mets\"ahovintie 114, 02540 Kylm\"al\"a, Finland; Max-Planck-Institut f\"ur Radioastronomie, Auf dem H\"ugel 69, 53121 Bonn, Germany}
\ead{tuomas.k.savolainen@aalto.fi}

%%%%%%%%%%%%%%%%%%%%%%%%%%%%%%%%%%%%%%%%%%%%%%%%%%%%%%%%%%%%

\author[s1]{Jose Luis G\'omez}%\fnref{footnote4}}
\address[s1]{Instituto de Astrof\'isica de Andaluc\'ia, CSIC,  Glorieta de la Astronom\'ia s/n, 18008 Granada, Spain}
%\fntext[footnote4]{Additional information about the co-authors}
\ead{jlgomez@iaa.es}

%%%%%%%%%%%%%%%%%%%%%%%%%%%%%%%%%%%%%%%%%%%%%%%%%%%%%%%%%%%%

\author[i3]{Andrei P. Lobanov}%\fnref{footnote4}}
%\address[g1]{Max-Planck-Institut f\"ur Radioastronomie, Auf dem H\"ugel 69, 53121 Bonn, Germany}
%\fntext[footnote4]{Additional information about the co-authors}
\ead{alobanov@mpifr-bonn.mpg.de}

%%%%%%%%%%%%%%%%%%%%%%%%%%%%%%%%%%%%%%%%%%%%%%%%%%%%%%%%%%%%

\author[a1,a2,i3]{Yuri~Y.~Kovalev}

\address[a1]{Astro Space Center of Lebedev Physical Institute, Profsoyuznaya~St.~84/32, 117997~Moscow, Russia}
\address[a2]{Institute of Physics and Technology,  Dolgoprudny, Institutsky per., 9, Moscow region, 141700, Russia}
%\address[a3]{Max-Planck-Institut f\"ur Radioastronomie, Auf dem H\"ugel 69, 53121 Bonn, Germany}

\ead{yyk@asc.rssi.ru }

%%%%%%%%%%%%%%%%%%%%%%%%%%%%%%%%%%%%%%%%%%%%%%%%%%%%%%%%%%%%

\author{\\on behalf of the \emph{RadioAstron} AGN imaging KSP teams}

%%%%%%%%%%%%%%%%%%%%%%%%%%%%%%%%%%%%%%%%%%%%%%%%%%%%%%%%%%%%
%%%%%%%%%%%%%%%%%%%%%%%%%%%%%%%%%%%%%%%%%%%%%%%%%%%%%%%%%%%%

\begin{abstract}
Imaging relativistic jets in active galactic nuclei (AGN) at angular resolution significantly surpassing that of the ground-based VLBI at centimetre wavelengths is one of the key science objectives of the \emph{RadioAstron} space-VLBI mission. There are three \emph{RadioAstron} imaging key science programs that target both nearby radio galaxies and blazars, with one of the programs specifically focusing on polarimetry of the jets. The first images from these programs reach angular resolution of a few tens of microarcseconds and reveal unprecedented details about the jet collimation profile,  magnetic field configuration, and Kelvin-Helmholtz instabilities along the flow in some of the most studied AGN (3C\,84, BL Lac, 3C\,273, S5\,0836+710). Here we give an overview of the goals and strategy of these three ongoing programs, highlight their early results, and discuss the challenges of space-VLBI imaging.

%Since 2013, three Key Science Programmes on Nearby AGN, Strong AGN, and AGN polarization, are being carried out with \emph{RadioAstron}, aiming at performing images and polarimetry of radio jets in AGN at the highest angular resolution available to date (tens of $\mu$as). First results from the different programmes show at unprecedented details the jet collimation profile, the magnetic field configuration, and the Kelvin-Helmholtz instabilities along the flow in some of the most studied AGN (3C\,84, BL Lac, 3C\,273, S5\,0836+710). In the following we give an overview on the goals and strategy of the three projects.
\end{abstract}


\end{frontmatter}

\parindent=0.5 cm

%%%%%%%%%%%%%%%%%%%%%%%%%%%%%%%%%%%%%%%%%%%%%%%%%%%%%%%%%%%%%%%%%%%%%%%%%%%%%

\section{Introduction}

Launched in 2011, \emph{RadioAstron} is a mission to realize a very long baseline interferometry (VLBI) array that combines ground radio telescopes with a 10-meter Space Radio Telescope (SRT) on a highly elliptical orbit around the Earth \citep{2013ARep...57..153K}. The antenna is equipped with four different on-board receivers, operating at 1.19$-$1.63\,cm (K-band), 6.2\,cm (C-band), 18\,cm (L-band), and 92\,cm (P-band). The apogee height of the orbit of Spektr-R spacecraft carrying the SRT is up to 360\,000\,km, which means that minimum spacing of interferometric fringes on the sky can be as small as 7\,$\mu$as at the \emph{RadioAstron}'s highest observing frequency of 22\,GHz. This makes \emph{RadioAstron} the highest angular resolution instrument in the history of astronomy and allows one to probe a previously unexplored parameter space for black hole powered jets in active galactic nuclei, pulsars, and galactic as well as extragalactic masers \citep{2017SoSyR..51..535K}.

At the beginning of \emph{RadioAstron} open science program in 2013, three key science programs (KSP; essentially large projects with legacy value) carrying out space-VLBI imaging of AGN were conceived to exploit the ultra-high angular resolution and polarimetry capabilities of the mission. The key scientific drivers of these programs were:

\begin{itemize}
    \item Resolving the jets in a few nearby AGN down to spatial scales of a few to a few hundred gravitational radii ($r_\mathrm{g}$) from the black hole (Nearby AGN KSP; PI: T.~Savolainen). Measuring the size, shape and internal structure of the region where the jet is accelerated and collimated provides a way to test the current models for jet formation in accreting black holes and answer questions like "do the current general relativistic magnetohydrodynamic simulations of jet formation \citep[e.g.,][]{2011MNRAS.418L..79T} capture all the relevant physics of the system" and "are the jets powered by accretion \citep{ 1982MNRAS.199..883B} or by rotational energy of the black hole itself \citep{1977MNRAS.179..433B}".      

    \item Transversely resolving the internal jet structure in powerful blazars and tracing shocks and plasma instabilities developing in the flow (Powerful AGN KSP; Co-PIs: A.~P.~Lobanov and M.~Perucho). Measuring the detailed morphology of the instability patterns and their comparison to earlier space-VLBI observations by the Japanese VSOP program \citep[e.g.,][]{2001Sci...294..128L} can help to distinguish between different plasma instabilities and consequently to constrain the physical conditions of the jet and the ambient medium. 

    \item Using ultra-high angular resolution multi-frequency polarization imaging to probe the jet magnetic field structure in or close to the jet acceleration and collimation zone (AGN polarization KSP; PI: J.~L.~G\'omez). In the current view of jet formation, the jets are highly magnetized with an ordered large-scale field near the launching site \citep{2011MNRAS.418L..79T, 2014Natur.510..126Z}, but the evolution of the magnetization and magnetic field structure further downstream is much less clear. Being able to transversely resolve the jets at multiple frequencies allows one to constrain the magnetic field structure by analyzing spatially resolved linear polarization spectra \citep[e.g.,][]{2013MNRAS.436.3341Z, 2016ApJ...817...96G}. 
\end{itemize} 

In this paper we give an overview of these three programs, their early results, and of the technical challenges of imaging observations with a space-VLBI satellite on a high orbit.

\section{Observations}

The three imaging programs have been running since 2013 and observations have been approved until 2019. The largest number of observed sources (ten; see Table~\ref{pol}) are in the polarization program, while the number of targets in the two other programs is more modest. Three nearby AGN were observed at multiple frequencies and epochs: 1.6\,GHz (M\,87 in 2014), 4.8\,GHz (M\,87 in 2014, 3C\,84 in 2013, Cen~A in 2014) and 22\,GHz (M\,87 in 2014 and 2018, 3C\,84 in 2013 and 2016, Cen~A in 2014). Furthermore, three powerful blazars (0836+710, 3C\,273, 3C\,345) were observed in 2013$-$2014 using the (non-polarimetric) dual-band mode of \emph{RadioAstron}. 

\begin{figure}
\begin{center}
\includegraphics[width=9cm]{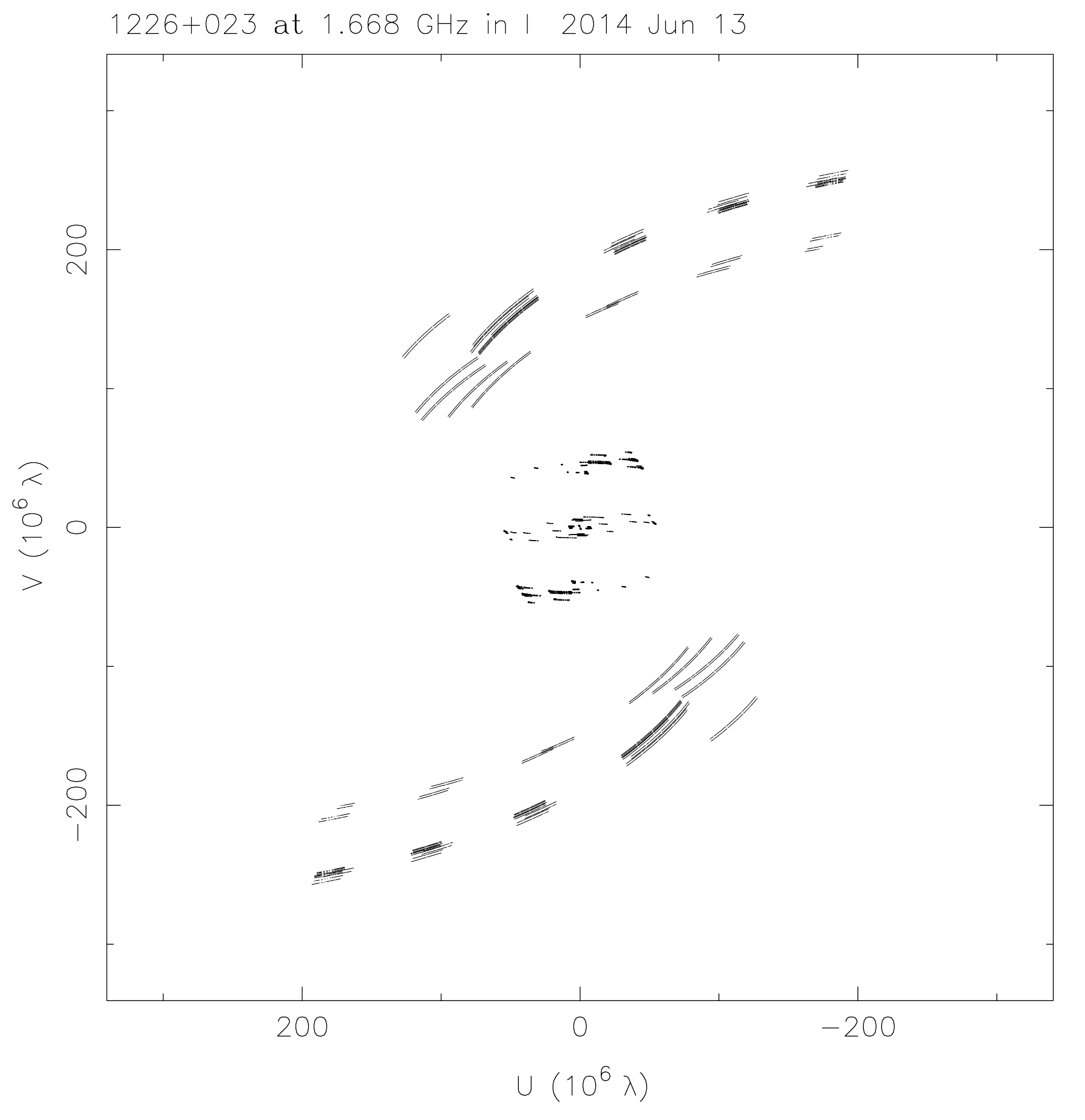}
\end{center}
\caption{An example $(u,v)$ coverage of \emph{RadioAstron} 3C\,273 observations at L-band, performed within the AGN polarization KSP ({\tt raks04f}). The central bulge, given by the ground array, traces about one Earth diameter, while wings are given by the \emph{RadioAstron} baselines. The maximum length of a space baseline in this figure is $\sim$4.5\,$D_\mathrm{Earth}$.}
\label{UV}
\end{figure}

All the imaging observations were carried out when the SRT was close to its perigee. Since the orbital velocity of the satellite is the highest near the perigee, it is possible to cover a wide range of ground-to-space baseline lengths, from less than one Earth diameter up to about ten Earth diameters, within a day. This is essential for the \emph{RadioAstron} imaging programs which are generally challenging due to the very sparse sampling of the ($u,v$) coverage for the ground-to-space baselines (see Fig. \ref{UV}). Continuous $\sim$24\,h long observations around the perigee passage require multiple tracking stations  \citep[Puschino near Moscow, Russia and Green Bank in West Virginia, USA;][]{2013ARep...57..153K,2014SPIE.9145E..0BF} and large ground radio telescopes over a wide range of longitudes in order to guarantee that there is always a common visibility between the SRT and a sensitive ground radio telescope. Sensitive baselines are crucial for finding the typically weak ground-to-space interferometric signal, i.e., fringes.   

%%%%%%%%%%%%%%%%%%%%%%%%%%%%%%%%%%%%%%%%%%%%%%%%%%%%%%%%%%%%

%%%%%%%%%%%%%%%%%%%%%%%%%%%%%%%%%%%%%%%%%%%%%%%%%%%%%%%%%%%%

Due to the these requirements, \emph{RadioAstron} imaging observations were typically supported on the ground by a global VLBI array including 15$-$30 antennas from the major astronomical cm-wavelength VLBI facilities: European VLBI Network (EVN) in Europe, Asia and Africa, Very Long Baseline Array (VLBA) in the United States, Long Baseline Array (LBA) in Australia, Kvazar Array in Russia, and Korean VLBI Network (KVN) in Korea. Furthermore, several large standalone radio observatories participated in the observations: Very Large Array (VLA), Green Bank Telescope and Arecibo in United States, Kalyazin in Russia, Usuda in Japan and Deep Space Network stations in Robledo, Spain and Tidbinbilla, Australia. The typical observing setup of the ground-based antennas was matched to the two 16\,MHz wide IFs of the SRT, but instead of the one-bit sampling of the \textit{RadioAstron} data, the ground array data were sampled with two bits. Therefore, in the typical observation \textit{RadioAstron} recorded data at the rate of 128\,Mbps, while the ground array recorded at 256\,Mbps. However, in polarimetric observations larger bandwidths were often used at the ground array in order to facilitate Faraday rotation measurements.

Since the imaging observations are limited to about 10\% of the full orbital period of 8.5 days and since the part of the sky that is visible at any given time is limited by a number of physical and operational constraints, scheduling of \emph{RadioAstron} imaging experiments is challenging. One may have to wait for the orbit to evolve for a year or more in order to find a date that provides sufficiently good ($u,v$) coverage for a particular target. This, in addition to the required substantial ground resources, typically limits the number of imaging experiment to a handful per year.

%Observations were performed in all the Announcement of Opportunity (AO) periods so far, supported by a global ground array including antennas from all the major interferometric facilities (European VLBI Network - EVN; Very Long Baseline Array - VLBA; Long Baseline Array - LBA; Kvazar array; Korean VLBI Array - KVN), as well as single dishes (Green Bank - US; Evpatoria, Kalyazin - Ukraine, Russian Federation; Arecibo - Puerto Rico; Usuda - Japan).

%%%%%%%%%%%%%%%%%%%%%%%%%%%%%%%%%%%%%%%%%%%%%%%%%%%%%%%%%%%%%%%%%%%%%%%%%%%%%

\section{Data processing}
The raw data from the different antennas participating in the imaging experiments were transferred to Max Planck Institute for Radio Astronomy (MPIfR) in Bonn, Germany, for correlation (see \citealt{2016ivs..conf..171M} for details on the correlator hardware and operation). Due to the nature of space-VLBI data, customized correlation and data reduction procedures are necessary, as explained below.

\subsection{Correlation of space-VLBI data}
A modified version of the {\tt DiFX} software correlator for VLBI \citep{2011PASP..123..275D}, able to combine ground and space antennas, was developed at the MPIfR. Among other features, this version can deal with data from an orbiting antenna, properly calculating data transmission delays as well as the general relativistic corrections for the delay model due to the Earth and Moon gravitational fields. Furthermore, a fringe-search window much wider than that provided by previous software can be used at the correlator stage, in order to look for signals with delay and delay-rate more than one order of magnitude larger than usually found for ground baselines. Indeed, uncertainties in the orbit reconstruction can easily lead to large discrepancies with respect to the expected fringe peak position. A complete description of the software is given in \cite{2016Galax...4...55B}. In case signal is not found at the longest space baselines, a best guess for delay and delay rate is extrapolated from the part of the experiment giving high-SNR fringes (typically when the SRT is close to the perigee), and a wider than usual correlation window (shorter integration time and smaller width of the spectral channels) is applied. This allows for further fringe search with post-processing software.

\subsection{Calibration and imaging}

Once data are correlated, a standard {\tt FITS} file is produced that can be imported for calibration with the {\tt PIMA}\footnote{http://astrogeo.org/pima/} \citep{2011AJ....142...35P} or {\tt AIPS}\footnote{http://www.aips.nrao.edu/index.shtml} \citep{2003ASSL..285..109G} interferometric data reduction and analysis packages and subsequently for imaging with e.g., the {\tt DIFMAP}\footnote{ftp://ftp.astro.caltech.edu/pub/difmap/difmap.html} \citep{1997ASPC..125...77S} software. 

One of the crucial steps in calibrating the space-VLBI data is searching for the often weak fringes on the noisy ground-to-space baselines. This requires techniques that are usually not necessary in the standard ground-VLBI fringe fitting. As mentioned before, delay and delay rate may significantly differ from the predicted values (up to $\sim2$\,$\mu$s and $7\times10^{-11}$\,s/s, respectively) due to inaccuracies in the SRT orbit prediction, which requires large fringe search windows. Secondly, the residual phase delay may have a significant second-order time dependence over the 10\,min scan lengths mostly this is an issue when the satellite is close to the perigee and its motion vector changes rapidly. This so-called \emph{acceleration} term can lead to decorrelation and in rare cases even to the loss of fringes. For many of the observed blazars the time variable delay rate is not a major problem, since they can often be detected with short enough integration time ($<1$\,min) so that average decorrelation due to the acceleration term remains small ($<1$\% for accelerations of $<$1 mHz/s, see \citealt{2016Galax...4...55B} for an example of this effect). However, the more resolved jets in the nearby AGN KSP require long integration times (often $\sim$10\,min) in fringe-fitting of the ground-to-space data. Fortunately, {\tt PIMA} is able to simultaneously solve for the acceleration term in addition to group delay and rate and it was used in the fringe search of the nearby AGN targets \citep[e.g., 3C\,84;][]{2018NatAs...2..472G}.

The bright target sources of the \emph{RadioAstron} AGN imaging programs can usually be easily detected on the ground-to-ground baselines. Therefore, we could in principle phase up the ground array and search for fringes between this coherently averaged signal and the SRT, which would lead to a reduction in the detection threshold. It turns out that the same reduction can be achieved by global fringe-fitting, if the ground-to-ground baselines are much more sensitive than the ground-to-space baselines \citep{KoganVLBAMemo13}. We have taken advantage of this by using an iterative fringe-fitting procedure. The ground-VLBI data are first processed in the standard manner, yielding a phase-calibrated data set and an image, which are used as inputs in the second round of global fringe-fitting. Fringes to the SRT are searched using multiple baseline combinations, so-called baseline stacking, which lowers the detection threshold. Furthermore, since the ground array data are already calibrated -- removing much of the short term atmospheric phase fluctuations\footnote{ We note that after self-calibration, the ground array data still contain phase fluctuations due to the atmosphere over the \emph{reference antenna}. However, this effect can often be minimized by selecting a well-behaved reference station.} -- it is possible to coherently average the data for the typical scan length of 10\,min\footnote{ Assuming that the acceleration term is small or already corrected.}. This procedure can enhance the sensitivity of space baselines, yielding fringe detections on baselines as long as 7$-$8 Earth diameters ($D_\mathrm{Earth}$) as in the case of the BL\,Lac observations in the AGN polarization program \citep{2016ApJ...817...96G} and 3C\,84 observations in the nearby AGN program \citep{2018NatAs...2..472G}. 

There are also a few important aspects of the \emph{RadioAstron} data that need to be taken into account when carrying out imaging and self-calibration steps. First, using natural weighting of the visibility data leads to a very small contribution to the image from the space baselines. This can be offset by weighting by the inverse of the local visibility sampling density in the ($u,v$) plane. This so-called \emph{uniform} weighting can be tuned by adjusting the bin size in the ($u,v$) plane over which the weights are determined. For space-VLBI, it is often beneficial to increase the bin size compared to what is used in the ground-only imaging. This is known as \emph{super-uniform} weighting and it enhances the contribution of the long ground-to-space baselines\footnote{ Sometimes the term "super-uniform" is also used to refer to \emph{not} scaling the weights by visibility amplitude errors.}. Secondly, if the signal-to-noise ratio on the ground-to-space baselines is low, it is dangerous to carry out phase self-calibration for the SRT using solution intervals as short as typically applied to the ground-array data ($\sim 10$\,s). Using a short phase self-calibration solution interval can lead to a generation of spurious flux from the noise on the ground-to-space baselines, if the baselines to the SRT are much longer than any of the ground-to-ground baselines \citep[for the general discussion of the effect, see][]{2008A&A...480..289M}. Luckily, the SRT is outside of the atmosphere and hence it suffers much less from the short time scale phase fluctuations than the ground-based telescopes. Hence, we used as long as 2\,min phase self-calibration solution interval for the SRT if the ground-to-space data had low SNR. Amplitude self-calibration of the SRT was typically limited to solving a single gain scaling factor for the whole observation.         

%Once \emph{a priori} amplitude calibration for \emph{RadioAstron} and ground antennas is applied, a first global fringe search for the ground array only is performed. This is particularly useful in order to later improve the signal to noise for space baselines, than can be challenging for resolved target structures. Once satisfactory solutions are found for the ground array, a second fringe search is performed applying baselines stacking, i.e. combining the ground baselines to improve the signal on space baselines with {\emph{RadioAstron}}. This procedure can significantly enhance the signal on space baselines, reaching 7-8 Earth Diameters (ED) as in the case of the BL Lac observations for the AGN polarization program \citep{2016ApJ...817...96G}. Imaging can be performed in {\tt DIFMAP}, and usually a ground-only image is fed back to {\tt AIPS} as a model to further improve the fringe search. The \emph{super-uniform} weighting scheme is often applied to enhance the contribution of the long ground-space baselines.

%%%%%%%%%%%%%%%%%%%%%%%%%%%%%%%%%%%%%%%%%%%%%%%%%%%%%%%%%%%%%%%%%%%%%%%%%%%%%

\section{Early results from the AGN imaging KSPs}
In the following, we give a brief summary of the work done since 2013 in the framework of the three KSPs, and the main results achieved to date.

%%%%%%%%%%%%%%%%%%%%%%%%%%%%%%%%%%%%%%%%%%%%%%%%%%%%%%%%%%%%

\begin{figure}
\begin{center}
\includegraphics[width=13cm]{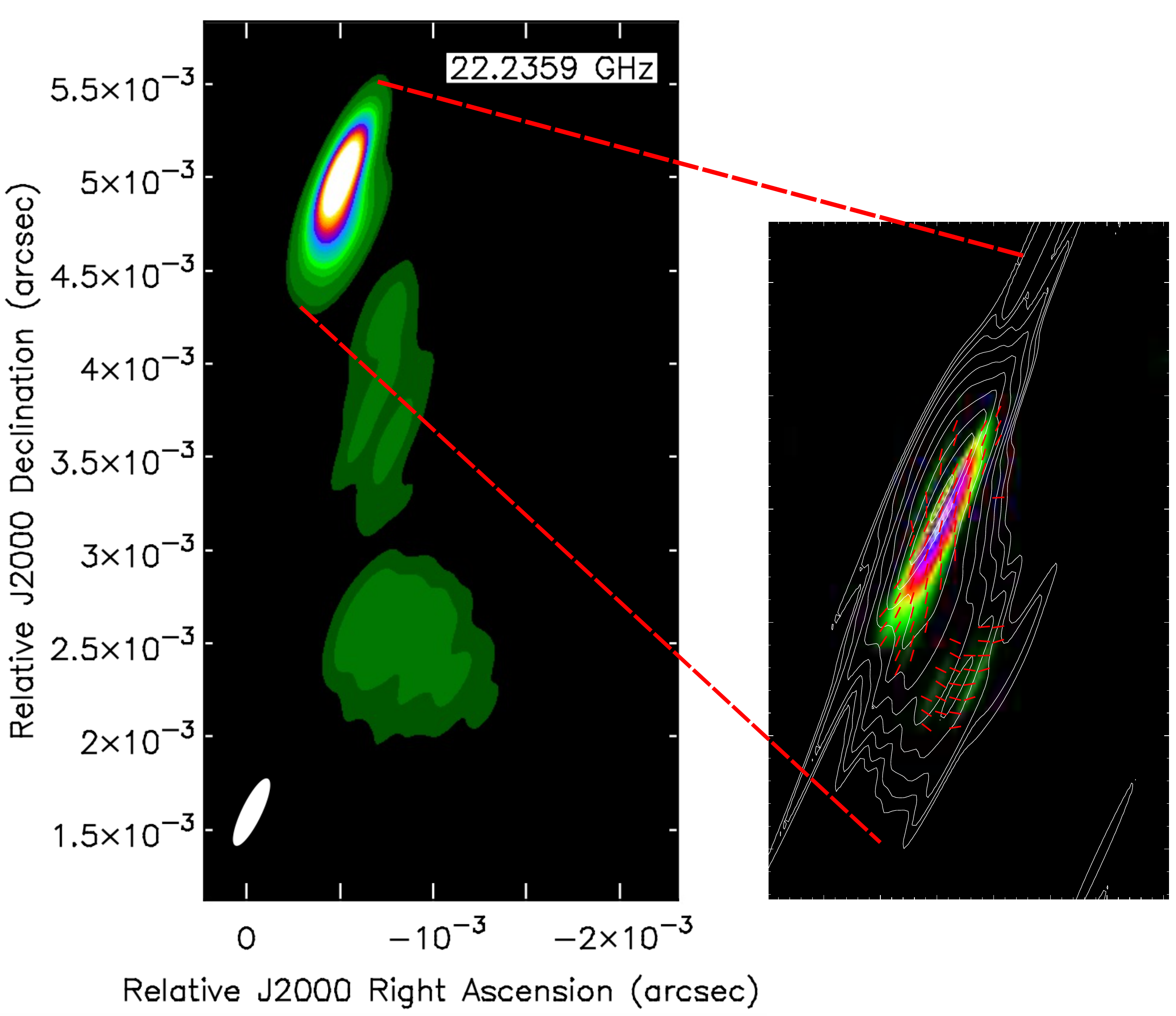}
\end{center}
\caption{Images of BL Lac at 22\,GHz from the AGN polarization KSP (\citealt{2016ApJ...817...96G}). Left panel: natural weighting, resulting in a synthesized beam of 390$\times$100\,$\mu$as; right panel: zoom on the core region in super-uniform weighting, reaching a resolution of 260$\times$21\,$\mu$as, with linearly polarized emission shown in colors, total intensity in contours, and Electric Vector Position Angle (EVPA) as red bars.}
\label{BLLac_maps}
\end{figure}

%%%%%%%%%%%%%%%%%%%%%%%%%%%%%%%%%%%%%%%%%%%%%%%%%%%%%%%%%%%%

\subsection{AGN Polarization KSP}

The AGN polarization KSP has been performing observations since the Early Science Period (ESP). With more than 20 experiments to date, it targeted several AGN with prominent polarimetric properties (see Table~\ref{pol}). During the ESP, the polarimetry capabilities of \emph{RadioAstron} were tested at L-band on 0642+499 \citep{2015A&A...583A.100L}, finding a stable instrumental polarization within 9\%, consistent between the two IFs. During the AO-1 period (July 2013 -- June 2014), the first 22\,GHz polarimetric observations on BL Lac were performed, giving fringes up to $\sim$8\,$D_\mathrm{Earth}$, and resulting in the highest angular resolution image to date (21\,$\mu$as, see Fig. \ref{BLLac_maps}, from \citealt{2016ApJ...817...96G}). Again, an instrumental polarization below 9\% was found, allowing one to use standard procedures to correct it. The resulting polarimetric image, combined with quasi-simultaneous ground-based polarimetric VLBI imaging at 15\,GHz and 43\,GHz, showed evidence for a large-scale helical magnetic field threading the jet. Moreover, evidence for emission from a (re-)collimation shock at 40\,$\mu$as upstream of the core was found. Further observations from the AO-1 period revealed a two orders of magnitude drop in the brightness temperature of 3C\,273 \citep{2017A&A...604A.111B} when compared to \emph{RadioAstron} observations performed with a similar configuration only one year before \citep{2016ApJ...820L...9K,2016ApJ...820L..10J}. Furthermore, 1.6\,GHz and 4.8\,GHz observations on the same target revealed a limb-brightened jet, with a sheath emission dominating at 1.6\,GHz, and a spine emission at 4.8\,GHz (Bruni et al.\ in prep.). Observations from the AO-2 period (July 2014 -- June 2015) of S5\,0716+71 probed the inner 100\,$\mu$as of the blazar jet, detecting polarized emission up to 5.6\,$D_\mathrm{Earth}$ space baselines, and a core brightness temperature exceeding the theoretical limits (see the contribution from Kravchenko et al.). 

Starting from 2017, close-in-time 86 and 230\,GHz observations at tens of $\mu$as resolution with the Global mm-VLBI Array and the Event Horizon Telescope \citep{2016Galax...4...54F} including the phased ALMA array have been providing unique data sets that will explore the jet structure and polarization properties of several AGN from 22\,GHz up to 300\,GHz.  

%%%%%%%%%%%%%%%%%%%%%%%%%%%%%%%%%%%%%%%%%%%%%%%%%%%%%%%%%%%%

\begin{table}
\caption{Details on observations preformed in the framework of the AGN polarization KSP.}
\scalebox{0.8}{
\begin{tabular}{lclccl}
\hline
Target & Date	& Exp. & Band &	Corr. &	Status and complementary obs.\\
\hline
0642+499 &	 03/2013 &	GK047	& L	 &  Yes	 & {\cite{2015A&A...583A.100L}}  \\
BL Lac	 &	09/2013 &	GA030A	& L	 &  Yes	 & Data analysis                                \\
BL Lac	 &	11/2013 &	GA030B	& K	 &  Yes	 & {\cite{2016ApJ...817...96G}}                 \\
3C273	 &	01/2014 &	GA030C	& K	 &  Yes	 & {\cite{2017A&A...604A.111B}}                 \\
3C273	 &	06/2014 &	GA030F	& L	 &  Yes	 & Bruni et al. (in prep.)                      \\
3C279	 &	03/2014 &	GA030D	& K	 &  Yes	 & Data analysis                                \\
OJ287	 &	04/2014 &	GA030E	& K	 &  Yes	 & G\'omez et al. (in prep.)                    \\
0716+714 &	 01/2015 &	GL041A	& K	 &  Yes	 & Kravchenko et al. (in prep.)                 \\
3C345	 &	03/2016 &	GG079A	& L	 &  Yes	 & Data analysis                                \\
OJ287	 &	04/2016 &	GG079B	& L	 &  Yes	 & Data analysis                                \\
OJ287	 &	04/2016 &	GG079C	& K	 &  No	 & Processing                                   \\
3C345	 &	  05/2016 &	GG079D	& K	 &  No	 & Processing                                   \\
3C454.3	 &	 10/2016 &	GG081A	& K	 &  No	 & Processing, GMVA              \\
CTA102	 &	10/2016 &	GG081B	& K	 &  No	 & Processing, GMVA                                         \\
OJ287	 &	03/2017	 &	GG081C	& K	 &  Yes	 & Processing, EHT+ALMA and GMVA+ALMA                       \\
BL Lac	 &	10/2017	 &	GG083A	& K	 &  No	 & Processing, GMVA                                         \\
3C279	 &	01/2018 &	GG083B	& K	 &  No	 & Processing, GMVA+ALMA                                    \\
3C120	 &	02/2018	 &	GG083C	& K	 &  No	 & Processing, GMVA                                         \\
3C273	 &	02/2018	 &	GG083D	& K	 &  No	 & Processing, GMVA+ALMA                                    \\
OJ287	 &	04/2018 &	GG083E	& K	 &  No	 & Processing, EHT+ALMA and GMVA+ALMA                       \\
\hline
\end{tabular}
}
\label{pol}
\end{table}

%%%%%%%%%%%%%%%%%%%%%%%%%%%%%%%%%%%%%%%%%%%%%%%%%%%%%%%%%%%%

\begin{figure}
\begin{center}
\includegraphics[width=12cm]{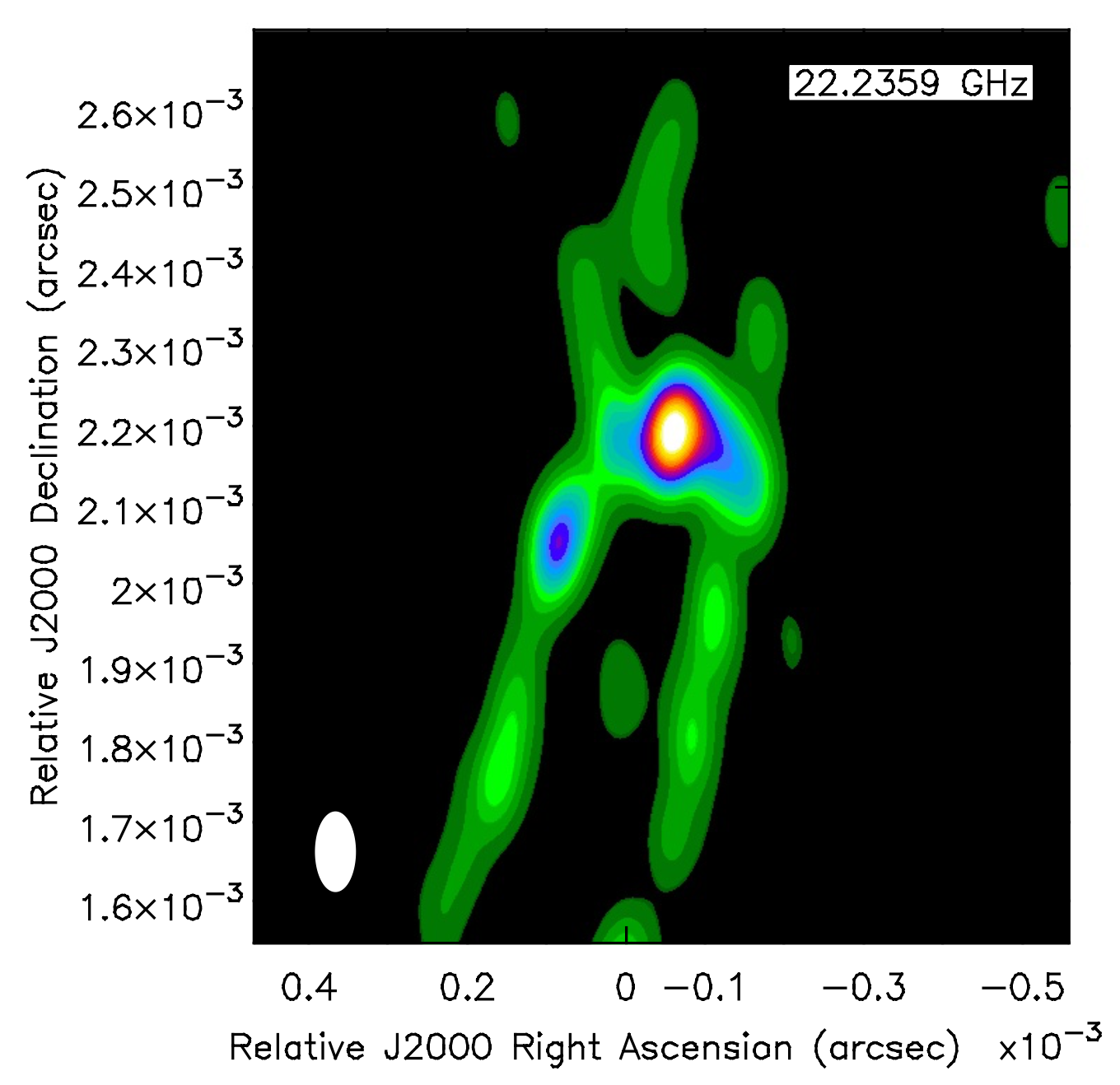}
\end{center}
\caption{Image of the nuclear region of 3C84 at 22 GHz, as seen by \emph{RadioAstron} at angular resolution of 100$\times$50\,$\mu$as \citep{2018NatAs...2..472G}. A large transverse jet radius of $\gtrsim$250\,$r_g$ at only 350\,$r_g$ from the core was measured. }
\label{3C84}
\end{figure}

%%%%%%%%%%%%%%%%%%%%%%%%%%%%%%%%%%%%%%%%%%%%%%%%%%%%%%%%%%%%

\subsection{Nearby AGN KSP}

Nearby AGN key science program is aimed at obtaining images with a \emph{spatial} resolution of a few to a few hundred gravitational radii in nearby radio galaxies 3C\,84 ($D_L = 75$\,Mpc), M\,87 ($D_L = 16$\,Mpc), and Cen~A ($D_L = 4$\,Mpc). All targets have been observed at multiple \emph{RadioAstron} bands and at multiple epochs. Ground-to-space fringes have been successfully found for 3C\,84 at 4.8 and 22\,GHz, and for M\,87 at 1.6, 4.8 and 22\,GHz. Two experiments on Cen~A at 4.8 and 22\,GHz did not yield ground-to-space fringes. 

The early dual-band observations of 3C\,84, the central galaxy of Perseus cluster, in 2013 resulted in fringe detections on ground-to-space baselines as long as 8\,$D_\mathrm{Earth}$ at both 4.8\,GHz (Savolainen et al.\ in prep.) and 22\,GHz \citep[][see also Fig. \ref{3C84}]{2018NatAs...2..472G}. With 27\,$\mu$as fringe spacing, the 22\,GHz \emph{RadioAstron} observations transversely resolved the edge-brightened jet of 3C\,84, and allowed measuring the jet collimation profile down to $\sim10^2$\,$r_g$ from the central black hole. The reconstructed image revealed an initially very broad jet with a transverse radius $\gtrsim250$\,$r_g$ at merely 350\,$r_g$ from the core, which implies that the jet is either rapidly laterally expanding on scales below $\sim10^2$\,$r_g$ \emph{or} the outer layer of the jet is launched from the accretion flow and not from the ergosphere of the spinning black hole. The measured collimation profile is almost cylindrical and it clearly differs from the parabolic profiles measured for M\,87 \citep{2018ApJ...868..146N} and Cyg~A \citep{2016A&A...585A..33B}. We believe that this likely due to the re-started nature of the jet in 3C\,84: the bright feature ejected from the VLBI core around 2003 could be a parsec scale analogue of a kiloparsec scale hot spot, i.e., the point where the jet rams into the external medium, slows down and dissipates a significant fraction of it energy \citep{2018NatAs...2..472G, 2017ApJ...849...52N}. Kiloparsec scale jets can collimate to an almost cylindrical shape before entering the hot spot due to the uniform pressure cavity that the jet has created around it \citep{1998MNRAS.297.1087K} and we may see the same effect at parsec scale in 3C\,84. The 4.8\,GHz \emph{RadioAstron} image shows low luminosity, cocoon-like emission around the re-started parsec-scale jets, which supports this explanation (Savolainen et al.\ in prep.)

In June 2014, there was an opportunity to obtain an exceptionally good ($u,v$) coverage in space for the nearby radio galaxy M\,87. The observations carried out at 1.6\,GHz together with 26 ground radio telescopes resulted in ground-to-space fringe detections up to 5\,$D_\mathrm{Earth}$. The reconstructed high-dynamic-range images have a beam size of 1$-$2\,mas depending on the applied weights. They allow detailed mapping of the rich internal structure of the 450\,mas long jet, which is resolved in transverse direction by 4$-$12 resolution elements and appears to contain helical threads or filaments (Savolainen et al.\ in prep.). Data processing and analysis of the other M\,87 observations, including the 22\,GHz polarimetric observation that was carried out in May 2018 -- close-in-time to the Event Horizon Telescope observations of M\,87 at 230\,GHz -- are still ongoing at the time of writing.

\subsection{Powerful AGN KSP}

First results from this programme have been submitted at the time of writing, presenting L, C, and K-band observations of S5 0836+710 (Vega-Garc\'ia et al.\ subm.). A detailed study of the Kelvin-Helmholtz instabilities along the jet, on angular scales between 0.02 and 200\,mas, was performed, resulting in an indication of helical instability modes. All observations were carried out during AO-1 and AO-2, including also 3C\,273, 3C\,345, and 4C+69.21. See the contribution from Zensus et al.\ for further details on the results of this KSP.

%%%%%%%%%%%%%%%%%%%%%%%%%%%%%%%%%%%%%%%%%%%%%%%%%%%%%%%%%%%%%%%%%%%%%%%%%%%%%

\section{Future prospects}

%It is foreseen that the nearby AGN and polarization KSPs will continue to propose further imaging blocks as the continuous evolution of the \emph{RadioAstron}'s orbit will keep bringing imaging opportunities for new AGN with suitable ground-to-space ($u,v$) coverage. Furthermore, the ability to observe a few well-selected sources close-in-time with \emph{RadioAstron} at 22\,GHz, 

 Recent inclusion of phased Atacama Large Millimeter/submillimeter Array (ALMA) in global VLBI networks at short mm-wavelengths now offers high sensitivity for tens of $\mu$as resolution imaging from the ground \citep{2018PASP..130a5002M}. With ALMA, the Global mm-VLBI Array at 86\,GHz and the Event Horizon Telescope at 230\,GHz give an opportunity to obtain multi-frequency polarization imaging at angular resolution comparable to the \emph{RadioAstron} one. Such data can be used to constrain the magnetic field structure close to the jet base via the analysis of spatially resolved polarization spectra. 

In the future, the \emph{Millimetron} space observatory (\citealt{2014PhyU...57.1199K}), led by the Astro Space Center of the Lebedev Physical Instityte, is planned to perform space-VLBI observations at mm-wavelengths. In the VLBI-mode the angular resolution of \emph{Millimetron} observations is in principle high enough to resolve black hole silhouettes at the centers of AGN up to a distance of $\sim$100\,Mpc, and probe at unprecedented resolution the matter outflows and magnetic field structure in the vicinity of the events horizon.

%%%%%%%%%%%%%%%%%%%%%%%%%%%%%%%%%%%%%%%%%%%%%%%%%%%%%%%%%%%%%%%%%%%%%%%%%%%%%

\section*{Acknowledgments}
GB acknowledges financial support under the INTEGRAL ASI-INAF agreement 2013-025-R.1.
TS was supported by the Academy of Finland projects 274477, 284495, and 312496.
The RadioAstron project is led by the Astro Space Center of the Lebedev Physical Institute of the Russian Academy of Sciences and the Lavochkin Scientific and Production Association under a contract with the State Space Corporation ROSCOSMOS, in collaboration with partner organizations in Russia and other countries. This research is based on observations processed at the Bonn Correlator, jointly operated by the Max Planck Institute for Radio Astronomy (MPIfR) and the Federal Agency for Cartography and Geodesy (BKG). The National Radio Astronomy Observatory is a facility of the National Science Foundation operated under cooperative agreement by Associated Universities, Inc. The European VLBI Network is a joint facility of independent European, African, Asian, and North American radio astronomy institutes.

%%%%%%%%%%%%%%%%%%%%%%%%%%%%%%%%%%%%%%%%%%%%%%%%%%%%%%%%%%%%
%%%%%%%%%%%%%%%%%%%%%%%%%%%%%%%%%%%%%%%%%%%%%%%%%%%%%%%%%%%%

%% Appendices
% The Appendices part is started with the command \appendix;
% appendix sections are then done as normal sections
% \appendix

%%%%%%%%%%%%%%%%%%%%%%%%%%%%%%%%%%%%%%%%%%%%%%%%%%%%%%%%%%%%

%\clearpage

%%%%%%%%%%%%%%%%%%%%%%%%%%%%%%%%%%%%%%%%%%%%%%%%%%%%%%%%%%%%
%%%%%%%%%%%%%%%%%%%%%%%%%%%%%%%%%%%%%%%%%%%%%%%%%%%%%%%%%%%%
\end{document}